\documentclass[a4paper,
              ]{jacow}
\makeatletter%
	\ifboolexpr{bool{xetex}}
	 {\renewcommand{\Gin@extensions}{.pdf,%
	                    .png,.jpg,.bmp,.pict,.tif,.psd,.mac,.sga,.tga,.gif,%
	                    .eps,.ps,%
	                    }}{}
\makeatother

\ifboolexpr{bool{xetex} or bool{luatex}} 
 {}                                      
 {\usepackage[utf8]{inputenc}}           

\usepackage[USenglish]{babel}			 

\usepackage[final]{pdfpages}
\usepackage{multirow}
\usepackage{ragged2e}

\ifboolexpr{bool{jacowbiblatex}}%
 {%
  \addbibresource{jacow-test.bib}
  \addbibresource{biblatex-examples.bib}
 }{}
\def\unsty{\,\hbox}

\begin{document}

\title{COMPENSATED PHASE JUMP AT TRANSITION CROSSING\protect\\IN THE CERN PS}

\author{H. Damerau, CERN, Geneva, Switzerland}
	
\maketitle

\begin{abstract}
The transition energy must be crossed in the CERN PS to accelerate proton and ion beams to flat-top energy. A phase jump of the accelerating RF voltage with respect to the phase of the bunches by about 180 degrees minus twice the synchronous phase is required at the instant of transition. This phase offset is injected into the beam phase loop which locks the phase of the vector sum of the RF voltage in the cavities to the beam. The polarities of the cavity return signal and of the stable phase programme are usually flipped at transition. However, both actions are difficult to perfectly synchronize in time, causing the beam phase loop to partially lock out and relock at the new stable phase. The resulting glitch can be avoided by well-controlled phase jumps applied to both, cavity drive and return signals simultaneously. This improved implementation of transition crossing makes it virtually transparent to the beam phase loop. The new scheme has been successfully tested with proton and ion beams, and it will become fully operational in the CERN PS after the long shutdown.
\end{abstract}

\section{INTRODUCTION}

Proton and ion beams must be accelerated through transition energy in the Proton Synchrotron (PS) at CERN. In the Super Proton Synchrotron (SPS) only LHC-type proton beams are injected just above transition, while fixed target proton and ion beams again cross this energy. At the transition energy the phase of the RF voltage with respect to the beam must be switched from the stable phase, $\phi_\mathrm{S}$ at the rising edge of the accelerating voltage to $180^0 - 2 \phi_\mathrm{S}$, the equivalent voltage at the bunch centre, but with inverted RF focussing.  The PS is equipped with pulsed quadrupole magnets which allow to move the transition energy rapidly through the energy of the beam. This so-called $\gamma_\mathrm{tr}$-jump scheme~\cite{sorenssen1975} ensures that the instant of transition crossing is well defined and avoids that the beam stays close to transition energy for too long.

The stable phase change must be implemented in the longitudinal beam control system to prevent the beam phase loop, which regulates relative phase between RF voltage and the bunches, from removing it after transition crossing. The beam phase loop in the PS is AC-coupled such that static or slow phase changes are not corrected, and it is operated in combination with the radial loop to adjust the average RF frequency during acceleration through transition energy. The polarity of this loop must also be flipped at transition in all cases. However, keeping in mind that the radial loop has a much smaller bandwidth than the beam phase loop, the corrections of both loops to the RF frequency are well decoupled.

With the conventional transition crossing scheme as detailed below, perturbations are observed on the beam phase loop. Due to various switching actions, the loop actually unlocks briefly at transition and relocks again at the new stable phase. Due to the frozen synchrotron motion the effect on the beam of this oscillation of the RF phase and frequency remains acceptable, but the power amplifiers driving the cavities must deliver the RF current to quickly change the phase of the cavity voltage.

To avoid this uncontrolled behaviour of the beam phase loop at transition crossing, a new scheme has been developed. It relies on changing the phase of the drive signals to the cavities while simultaneously undoing the phase shift on the cavity return signals. When both phase changes are executed in a well controlled way, transition crossing becomes almost transparent to the phase loop, hence avoiding any transient conditions.

The compensated phase jump at transition crossing has been successfully tested with low and high intensity beams in the PS before the Long Shutdown 2 (LS2). It will become operational with the restart of the PS in 2020.

\section{CONVENTIONAL TRANSITION CROSSING SCHEME}

\begin{figure}[!tbh]
	\centering
	\includegraphics*[width=0.6\columnwidth,bb=145 265 450 575]{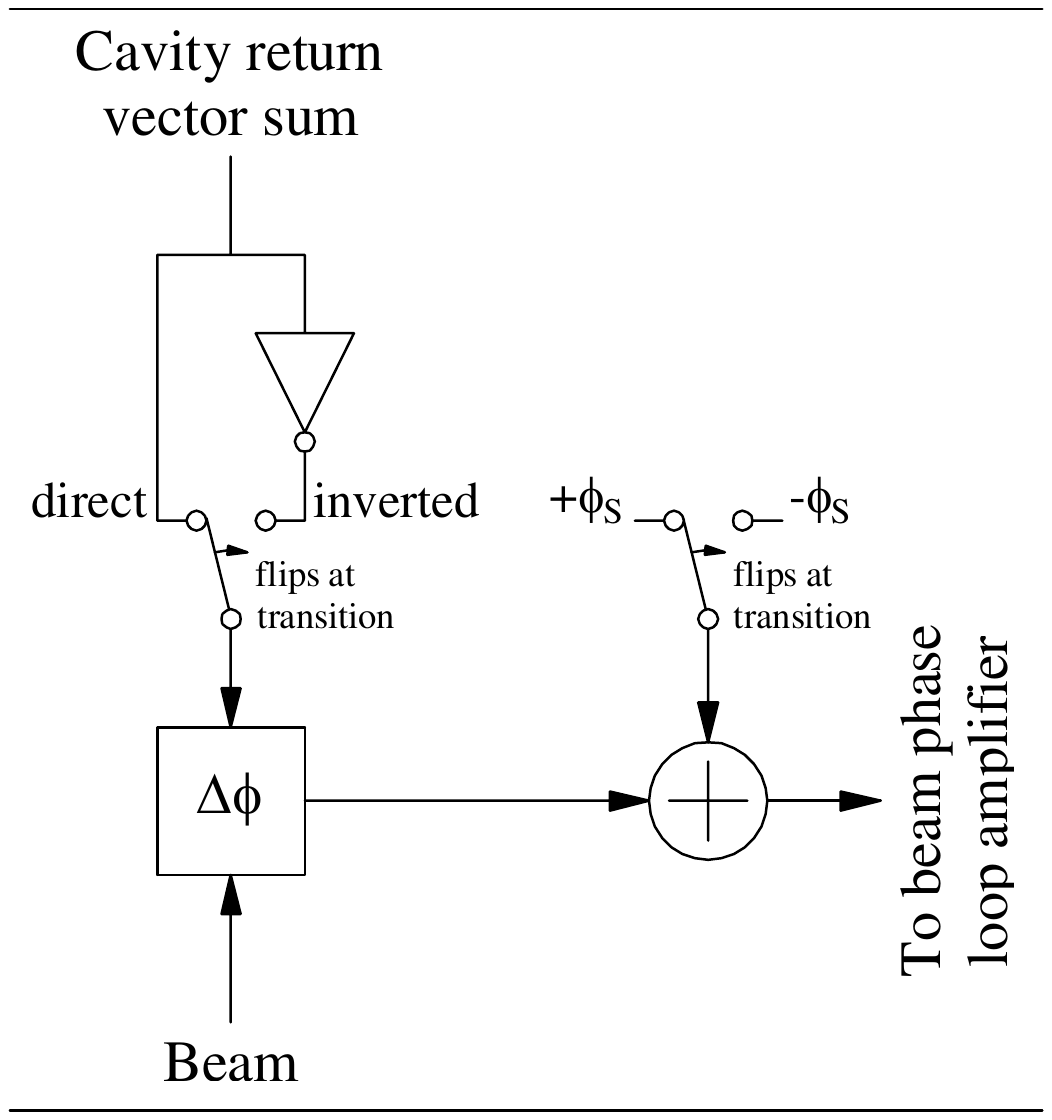}	
	\caption{Implementation of the phase jump at transition for the beam phase loop in the conventional scheme. The trigger for both switches is adjusted to result in a combined phase shift of $180^0 - 2 \phi_\mathrm{S}$.}
	\label{figSchematicsConventionalTransitionCrossing}
\end{figure}

With the conventional transition crossing scheme, the set-point change of the beam phase loop by $\Delta \phi_\mathrm{tr} = 180^0 - 2 \phi_\mathrm{S}$ is generated by two separate actions~(Fig.~\ref{figSchematicsConventionalTransitionCrossing}).
Firstly, the sign of the vector sum of the cavity return signals is inverted, equivalent to a de-phasing of $180^0$. Secondly, the sign of the stable phase program is switched from $\phi_\mathrm{S}$ to $-\phi_\mathrm{S}$ contributing the remaining $2 \phi_\mathrm{S}$ to $\Delta \phi_\mathrm{tr}$.

\begin{figure}[!tbh]
	\centering
	\includegraphics*[width=0.8\columnwidth]{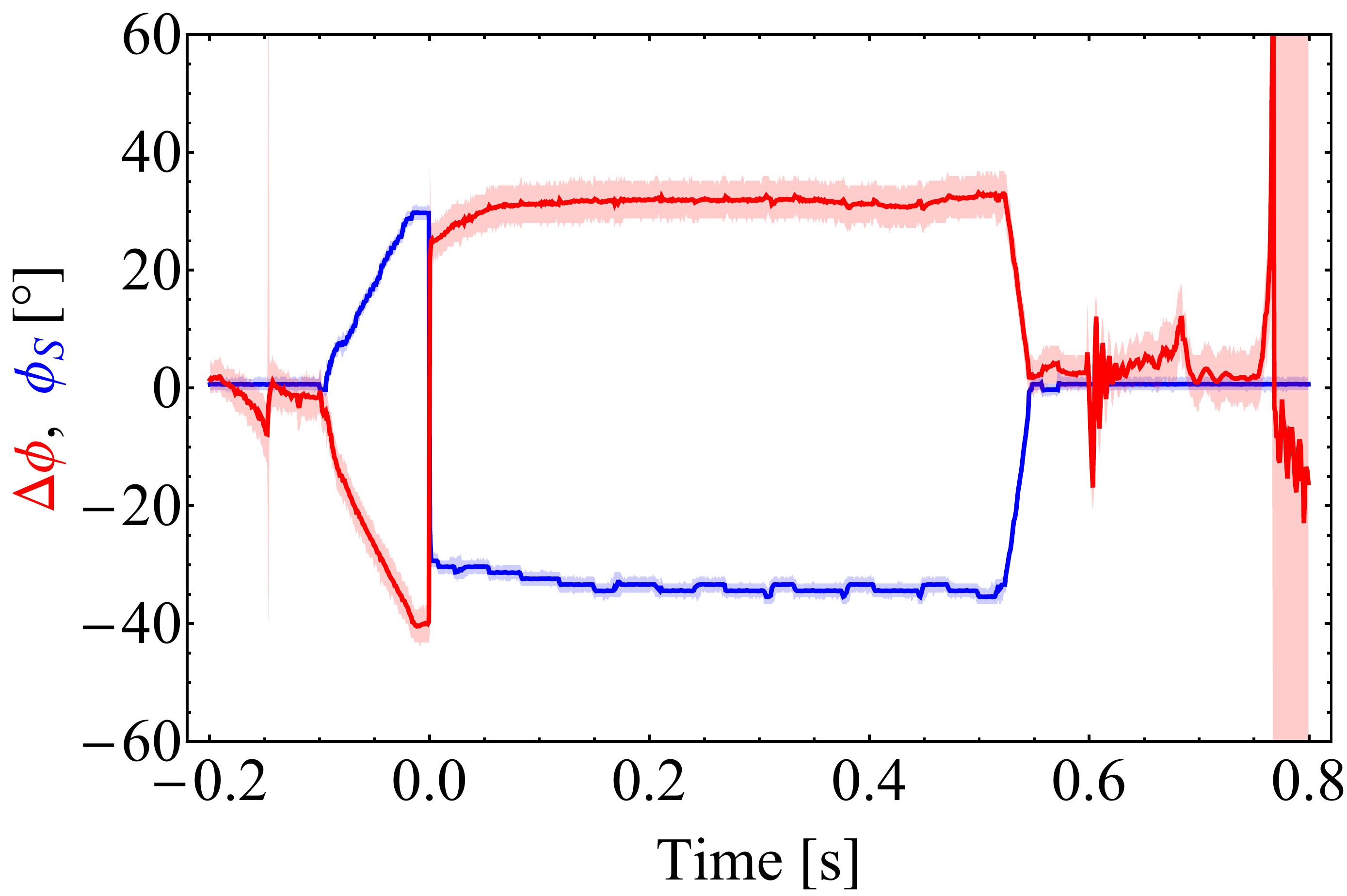}	
	\caption{Measured phase between beam and vector sum of cavity return signals during a proton cycle with LHC-type multi-bunch beam. The time reference is at transition crossing. The spike at about $150\unsty{ms}$ before transition crossing is caused by a glitch when the harmonic of the phase loop is switched from harmonic $h=7$ to $h=21$. The quantization noise on the stable phase programme is caused by insufficient resolution of the computation from ramp rate and total RF voltage in hardware.}
	\label{figPhaseStablePhaseFullCycleStandardTransition}
\end{figure}

The phase between the beam and the vector sum of the cavity return signals during the cycle is plotted in Fig.~\ref{figPhaseStablePhaseFullCycleStandardTransition}, together with the stable phase program in the case of conventional transition crossing. 
At constant energy below transition~(before $-0.1\unsty{s}$, $E_\mathrm{kin} = 2.5\unsty{GeV}$), the phase between beam and RF voltage is indeed close to zero. At the flat-top above transition~(after $0.55\unsty{s}$, $E_\mathrm{tot} = 26\unsty{GeV}$), a measured phase of zero corresponds to $180^0$ due to the inversion of the cavity return signal at transition.

\begin{figure}[!tbh]
	\centering
	\includegraphics*[width=0.8\columnwidth]{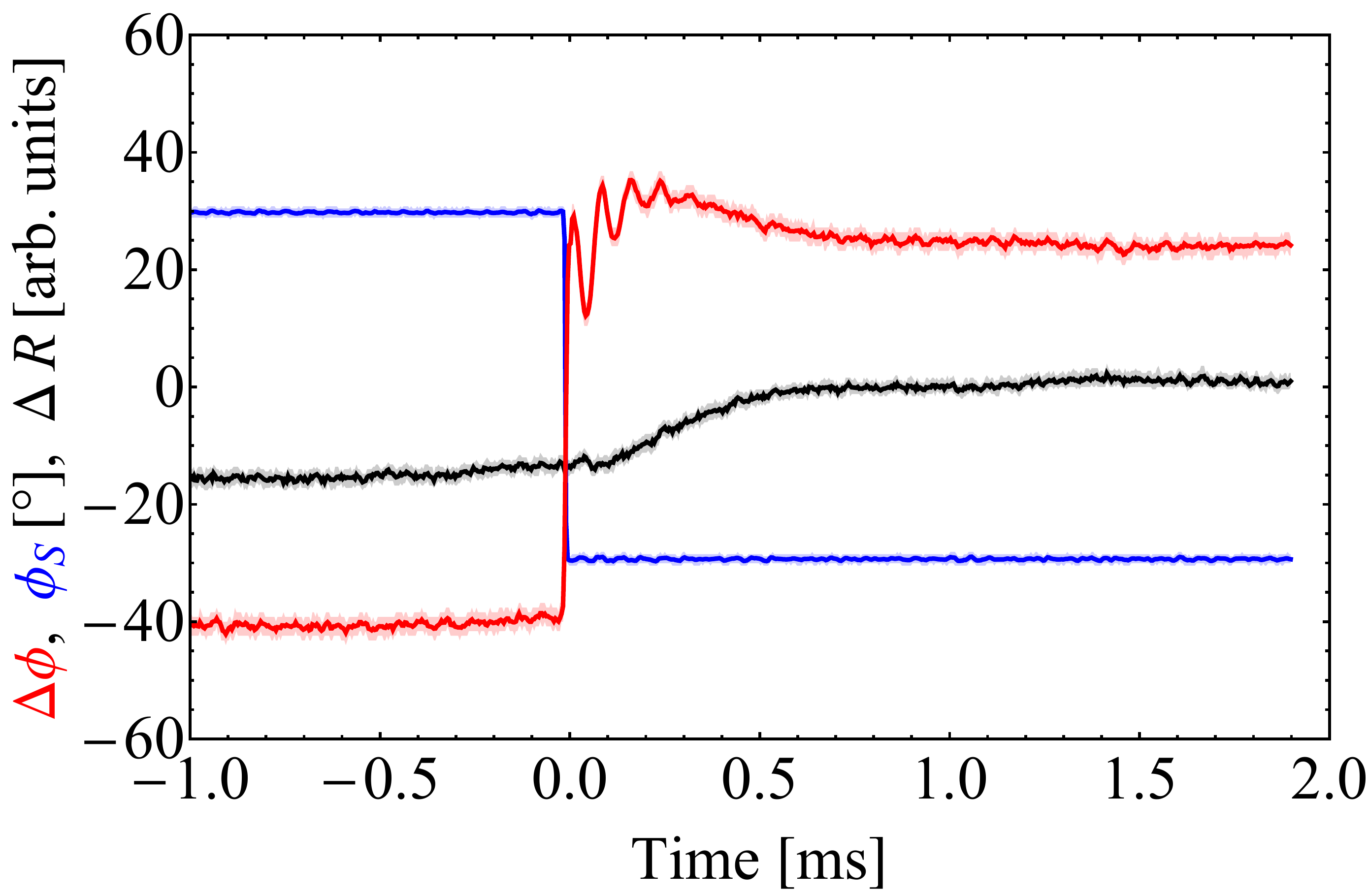}	
	\caption{Zoom of measured phase, $\phi$ (red), stable phase program, $\phi_\mathrm{S}$ (blue) and radial position offset, $\Delta R$ (black) around transition crossing.}
	\label{figPhaseStablePhaseStandardTransitionZoom} 
\end{figure}

The sign inversions of cavity return phase and stable phase are difficult to synchronize perfectly, keeping in mind that the former switches an RF signal while the latter is added to the phase between two RF signals with moderate bandwidth. Figure~\ref{figPhaseStablePhaseStandardTransitionZoom} illustrates the close view of the signals around transition crossing. 
Even under optimized conditions the phase of the RF voltage oscillates around the reference phase defined by the beam until the phase loop fully locks after only $400\unsty{$\mu$s}$ (almost 200~turns). This may also trigger oscillations of the bunch. The amplitude and duration of the oscillation are very sensitive to the gain of the beam phase loop, as well as to the relative alignment between cavity return and stable phase sign changes in terms of time and duration. In the worst case the phase loop may relock on the wrong RF period such that the phase of the distributed revolution frequency clock intermittently changes with respect to the position of bunches. 

\begin{figure}[!tbh]
	\centering
	\includegraphics*[width=0.75\columnwidth]{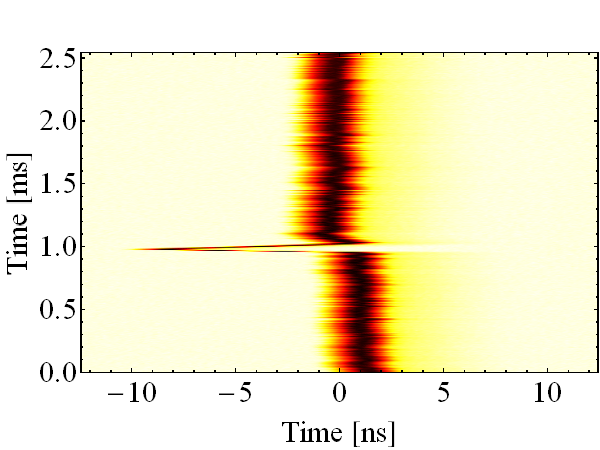}	
	\caption{Evolution of the bunch profile at transition crossing of a $^{208}$Pb$^{54+}$ ion bunch. A profile is recorded every 3 turns, triggered by the distributed revolution frequency~(Fig.~\ref{figSchematicsCompensatedTransitionCrossingOverview}). The period of the RF frequency, $1/f_\mathrm{RF}$ is $102.5\unsty{ns}$.}
	\label{figMountainRangeHighResolutionIonCrossing}
\end{figure}

A mountain range plot of a $^{208}$Pb$^{54+}$ ion bunch at transition crossing using the conventional crossing scheme is shown in Fig.~\ref{figMountainRangeHighResolutionIonCrossing}.
The glitch of the bunch arrival time is due to the distributed revolution frequency clock moving with respect to the bunch when the phase loop relocks after transition. It is important to note that a phase offset of $180^0-2\phi_\mathrm{S}$ is introduced in the drive signals to the cavities to align the bunches in time with respect to the revolution frequency clock.

\section{COMPENSATED PHASE JUMP}

The implementation of the compensated phase jump at transition crossing relies on the flexible multi-harmonic RF sources~\cite{damerau2017} driving the accelerating cavities in the PS, as well as on a universal cavity return based on the same sources. The so-called multi-harmonic~(MHS) RF sources generate beam synchronous signals at any harmonic number, which may even change through the acceleration cycle. Besides a slow control of the output phase required for RF manipulations~\cite{damerau2013,damerau2018}, they allow programmable phase jumps triggered by timings with a resolution down to a single clock cycle, $T_\mathrm{clock} = T_\mathrm{rev}/256$. This is achieved by re-programming the phase register of the DDS-like MHS at every clock cycle. As a result, the output phase can, e.g. be linearly swept at transition crossing by $180^0 - 2 \phi_\mathrm{S}$ in exactly one revolution period, corresponding to 256~clock cycles. The optimum duration of the phase flip is adjusted to reduce the transient for the power amplifier.

\subsection{Implementation}

\begin{figure}[!tbh]
	\centering
	\includegraphics*[width=0.7\columnwidth,bb=122 83 472 760]{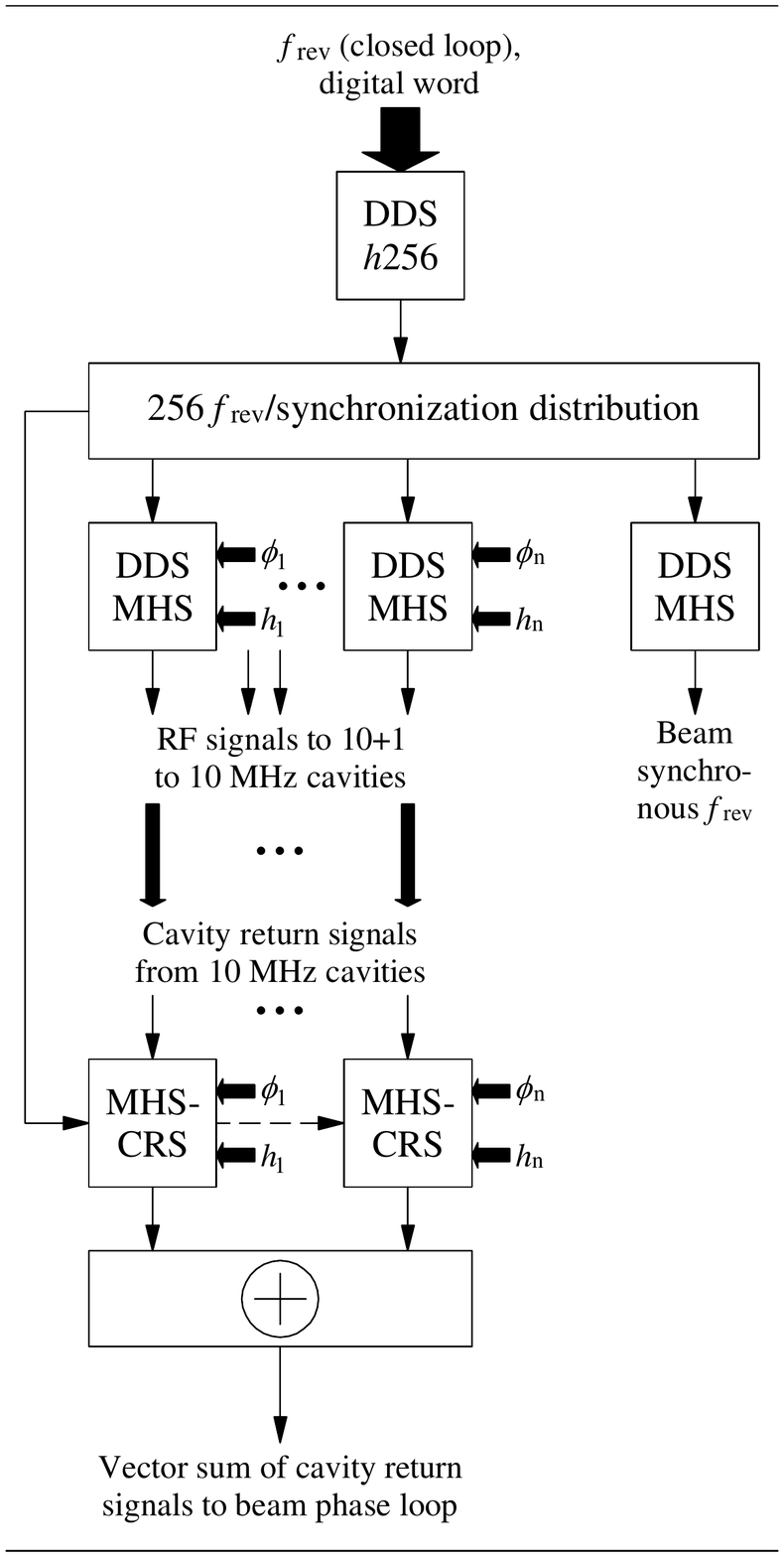}	
	\caption{Simplified diagram of the generation of the drive signals to the eleven main accelerating cavities and the signal processing to recombine the gap return signals of the individual cavities to the vector sum for the beam phase loop.}
	\label{figSchematicsCompensatedTransitionCrossingOverview}
\end{figure}

The universal cavity return de-phases the gap return signals from individual cavities to compensate for the azimuthal position of the corresponding cavity in the ring at any harmonic number~\cite{damerau2017} and sums them up. It is important to point out that this de-phasing undoes the phase offsets of $\phi = h \Phi$ applied to the drive signals of the cavites to operate them in phase with respect to the circulating beam; $\Phi$ is the azimuthal position of the cavity in the ring. Combining the de-phased cavity return signals forms the vector sum. Its phase is compared to the beam phase as an input to the phase loop (Fig.~\ref{figSchematicsCompensatedTransitionCrossingOverview}).
All MHS and MHS cavity return sum receivers~(MHS-CRS) are kept beam synchronous using a common clock at $256f_\mathrm{rev}$ and a synchronization pulse distributed during each cycle. The beam synchronous revolution frequency clock is generated in the same way as the drive signals to the cavities, using an MHS permanently set to $h=1$.

\begin{figure}[!tbh]
	\centering
	\includegraphics*[width=0.6\columnwidth,bb=162 115 433 725]{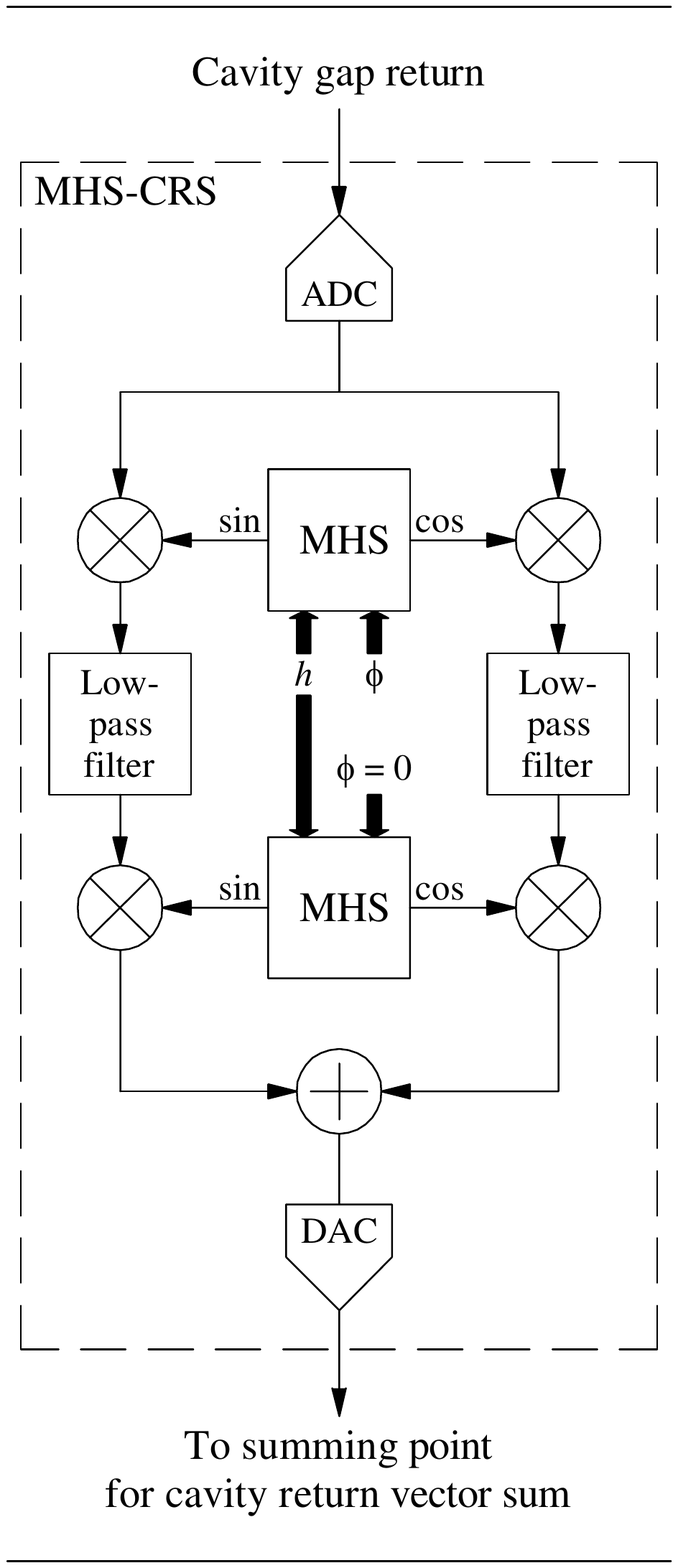}	
	\caption{Signal processing applied to the cavity return signals (MHS-CRS) to remove the de-phasing due to the azimuthal position of the cavity in the ring and to inject the phase jump of programmable amplitude and duration around transition crossing.}
	\label{figSchematicsCompensatedTransitionCrossingSingleChannel}
\end{figure}

As each cavity is driven by an MHS, with its gap voltage signal treated by an MHS-based signal processing, it becomes straightforward to inject a programmable phase jump at transition into the drive signal of each cavity and to remove the phase jump from the cavity return signal by introducing exactly the opposite phase offset to the cavity return signal. Hence the phase between beam and RF voltage moves without affecting any other signal. The signal processing for a single cavity return signal is sketched in Fig.~\ref{figSchematicsCompensatedTransitionCrossingSingleChannel}. 
Following direct digitization, the cavity return signal is digitally down-converted to baseband. The phase is controlled by acting on the phase $\phi$ of the MHS serving as a digital local oscillator. This phase contains the correction of the azimuthal position of the cavity in the ring to correct for the time of flight of the beam, as well as the programmed fast phase changes for transition crossing. The filtered baseband signals are then up-converted again to regenerate the de-phased input signal. Conceptually, the signal processing represents a narrow-band phase shifter, fully programmable in harmonic number and phase like the MHS generating the drive signals to the cavities. No additional dedicated hardware other than the trigger timing is hence required to implement the additional phase jump at transition.

Neglecting the beam induced voltage, the phase change of beam and RF voltage at transition crossing becomes transparent to the beam phase loop. To achieve full compensation the propagation delay through cables, power amplifier and cavity must also be taken into account, seen that the jump duration is with a few microseconds of the same order of magnitude. Consequently, the trigger of the phase jump to the MHS-CRS is delayed by this propagation delay.

\subsection{Results with beam}

\begin{figure}[!tbh]
	\centering
	\includegraphics*[width=0.8\columnwidth]{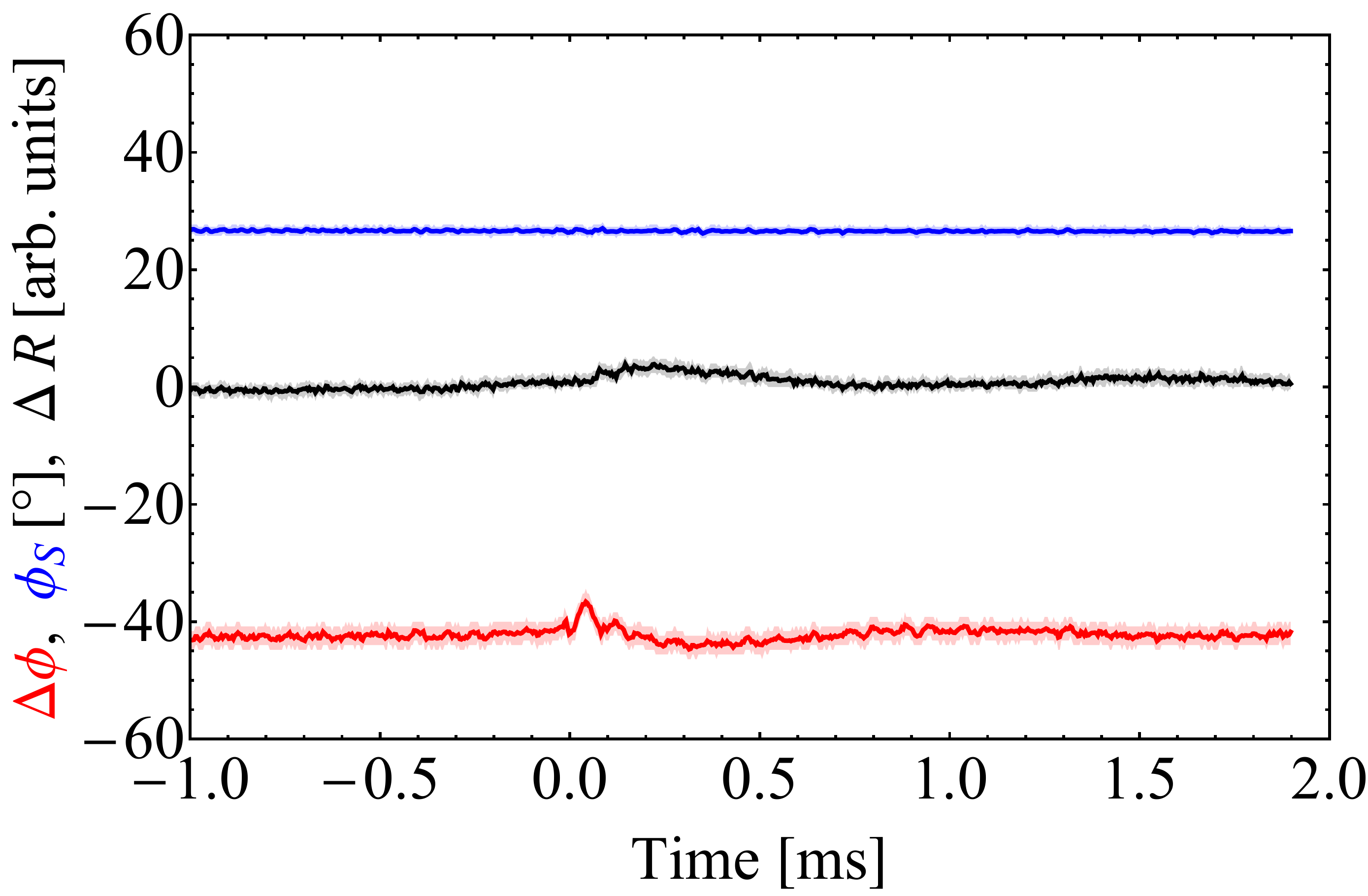}	
	\caption{Zoom of measured phase, $\phi$ (red), stable phase program, $\phi_\mathrm{S}$ (blue) and radial position offset, $\Delta R$ (black) around transition crossing with the new scheme. The beam conditions are the same as for Fig.~\ref{figPhaseStablePhaseStandardTransitionZoom}.}
	\label{figPhaseStablePhaseFullCycleCompensatedTransitionZoom}
\end{figure}

The measured loop signals at transition crossing with the compensated scheme are plotted in Fig.~\ref{figPhaseStablePhaseFullCycleCompensatedTransitionZoom}.
Since all phase changes are compensated by the MHS-CRS, the stable phase program does not change. It just represents the absolute phase offset with respect to the rising or falling zero crossing of the RF voltage. The phase difference between beam and RF, as well as the radial position offset only indicate very small residual perturbations. The transition crossing has become almost transparent to beam phase and radial loops.

Although the transition energy is passed virtually without any transient the absolute value of the phase measured between beam and RF voltage stays around $-\phi_\mathrm{S}$, instead of $+\phi_\mathrm{S}$ with the conventional crossing scheme. During acceleration this is not an issue, but it prevents the phase measurement to return to zero at the flat-top which is a necessary constraint to change the phase loop harmonic during RF manipulations.

\begin{figure}[!tbh]
	\centering
	\includegraphics*[width=0.8\columnwidth]{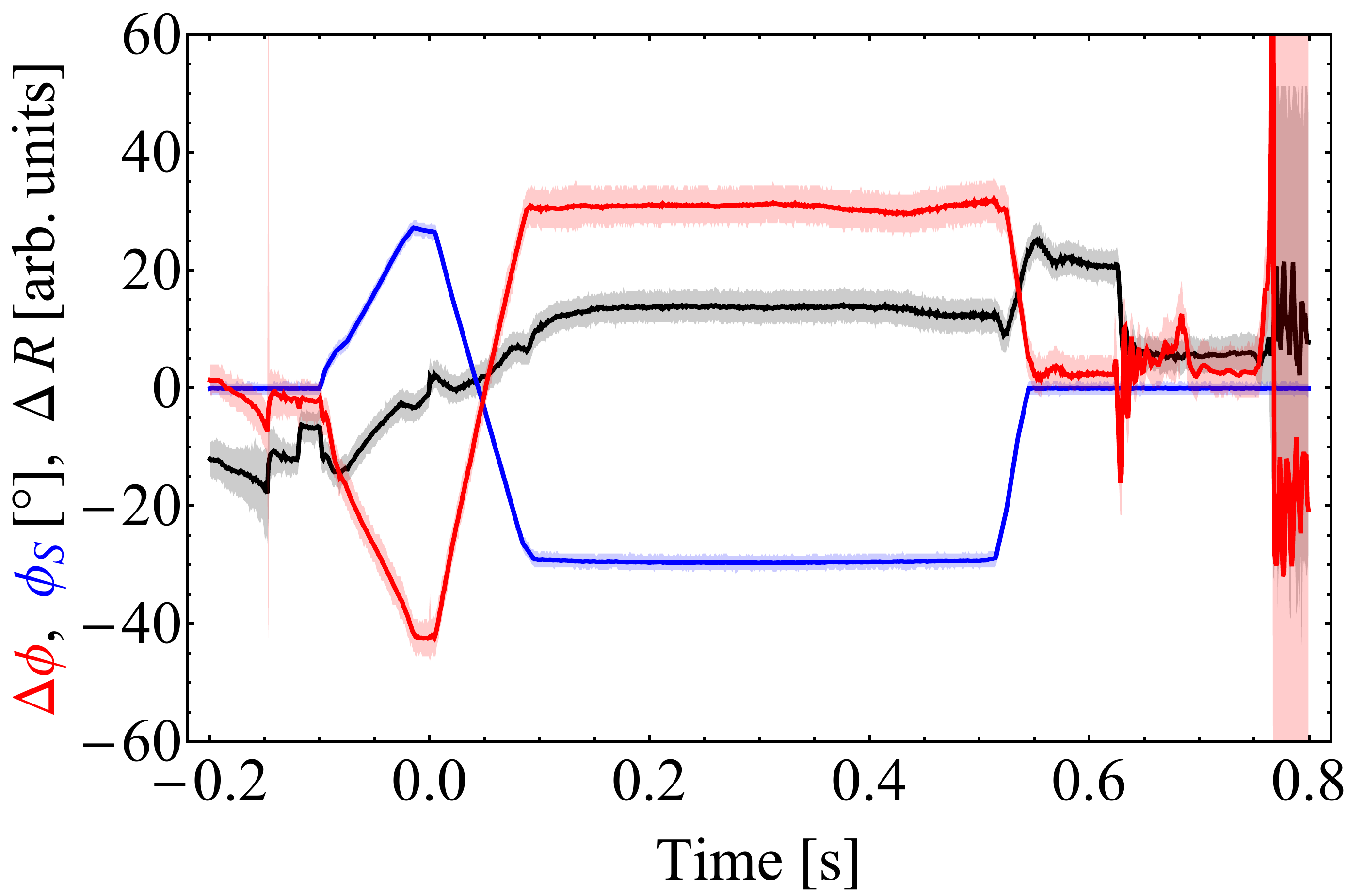}	
	\caption{Beam phase loop (red), radial position offset (black) and stable phase program during the cycle with the compensated transition crossing scheme. Identical conditions as for the measurements Fig.~\ref{figPhaseStablePhaseFullCycleStandardTransition} otherwise.}
	\label{figPhaseStablePhaseCompensatedTransition}
\end{figure}

Once the transition crossing is passed, the phase offset of the beam phase loop must hence be changed by slowly moving the de-phasing of the cavity vector sum signal by $2\phi_\mathrm{S}$~(Fig.~\ref{figPhaseStablePhaseCompensatedTransition}), typically on the time scale of $100\unsty{ms}$.
The stable phase program is simultaneously decreased from $\phi_\mathrm{S}$ to $-\phi_\mathrm{S}$. Both actions would ideally compensate each other such that the phase loop would not be affected at all. Tiny perturbations in the sub-millimetre range are nonetheless visible on the radial position at beginning and end of the stable phase correction, but without measurable impact on beam quality.

\begin{figure}[!tbh]
	\centering
	\includegraphics*[width=\columnwidth]{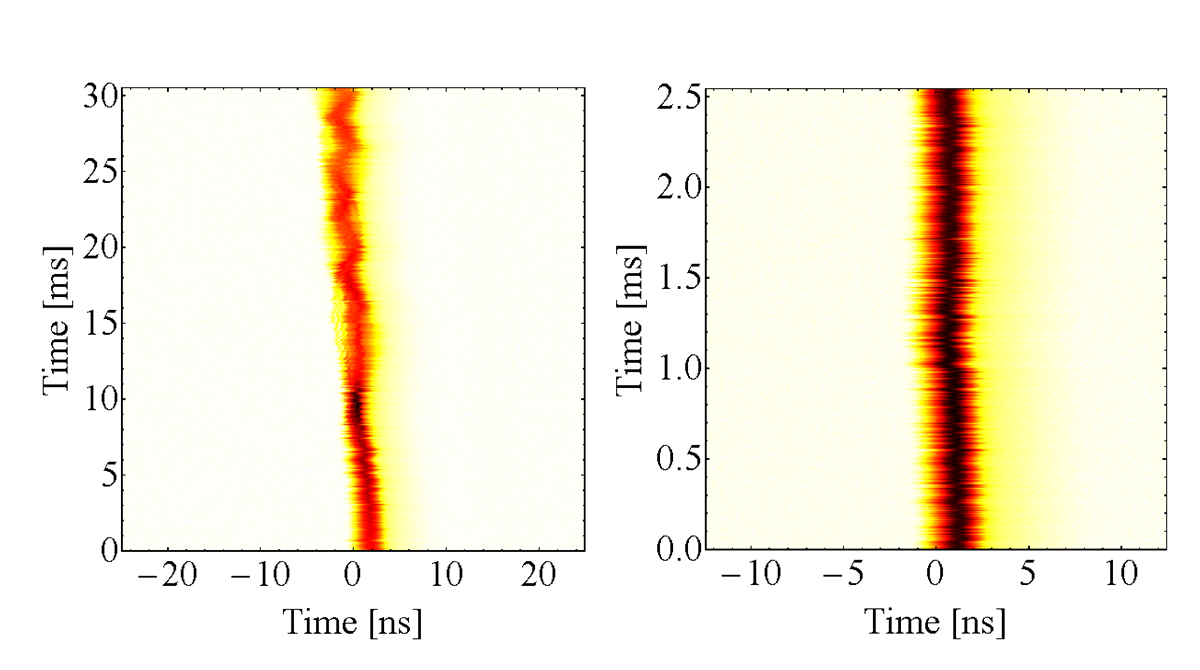}	
	\caption{Mountain range plot of an $^{208}$Pb$^{54+}$ ion bunch passing through transition energy with the new scheme on a timescale of $30\unsty{ms}$ (left) and zoomed closely around the phase flip (right). The bunch gets longitudinally unstable some $5\unsty{ms}$ after transition crossing which is due to intensity effects and unrelated to the transition crossing scheme.}
	\label{figMountainRangeIonsLargeTimeSpanAndZoom}
\end{figure}

Figure~\ref{figMountainRangeIonsLargeTimeSpanAndZoom} illustrates the quality of transition crossing with the new compensated crossing scheme. 
Since the bunch does not move with respect to the distributed revolution frequency clock, no glitch is visible and the bunch just shortens when approaching transition at stretches again thereafter. Even zooming closer around the phase jump (right) does not reveal any glitch. 

\section{CONCLUSIONS}

A compensated transition crossing scheme has been implemented in the PS low-level RF system and commissioned with beam. It removes the glitch of the beam phase loop at transition by changing the phase of cavity drive and return signals simultaneously such that the phase jump becomes transparent to the beam control loops. Despite from being essentially glitch free, the new scheme offers several advantages compared to the conventional scheme used at CERN before. Since the generation and vector summation of the cavity signals has been upgraded anyway to MHS-based systems, no additional hardware is required for the compensated transition crossing. The controlled, programmable phase jumps of the MHS allow optimizing the jump duration to facilitate the phase flip for the RF power systems. Furthermore, no fast change of the stable phase programme is required with the compensated crossing scheme, which significantly simplifies its generation. The new transition crossing scheme has been fully validated with all relevant proton and ion beams in the PS during the 2018 run. It will be the default configuration for the restart after LS2.

\section{ACKNOWLEDGEMENTS}

The author is grateful to the PS operations teams for their support of tests.

	\AtEndDocument{\par\leavevmode}

\end{document}